# Ternary Nb$_3$Sn superconductors with artificial pinning centers and high upper critical fields

X Xu[1], J Rochester[2], X Peng[3], M Sumption[2] and M Tomsic[3]

[1] Fermi National Accelerator Laboratory, Batavia, IL 60510, U.S.A

[2] Department of MSE, the Ohio State University, Columbus, OH 43210, U.S.A

[3] Hyper Tech Research Incorporated, 539 Industrial Mile Road, Columbus, OH 43228, U.S.A

E-mail: xxu@fnal.gov

**Abstract**

In this letter we demonstrate the development of ternary Nb$_3$Sn multifilamentary conductors with artificial pinning centers (APC) which achieve high critical fields. These recently-developed conductors were tested in a 31 T magnet, and the results showed that their upper critical field ($B_{c2}$) values at 4.2 K are 27-28 T, and irreversible field ($B_{irr}$) values are above 26 T, values similar to or higher than those of best RRP conductors. The non-Cu $J_c$ has been brought to nearly 1200 A/mm$^2$ at 16 T and 4.2 K, comparable to RRP, in spite of the fact that the fine-grain Nb$_3$Sn fractions in filaments are still low (20-30%) and the grain sizes are still not fully refined (70-80 nm) due to conductor designs and heat treatments that are not yet optimized. The Nb$_3$Sn layer $J_c$ at 4.2 K, 16 T is 4710 A/mm$^2$ for the APC wire with 1%Zr, about 2.5 times higher than RRP conductors, in spite of the fact that its grain size is not yet fully refined due to insufficient oxygen and unoptimized heat treatment. An analysis is presented about the non-Cu $J_c$ that can be achieved by further optimizing the APC conductors and their heat treatments.

# 1. Introduction

Several next-generation proton-proton circular colliders have been proposed to succeed the Large Hadron Collider (LHC), including the Future Circular Collider (FCC) and the High-energy LHC (HE-LHC), for both of which 16 T magnets based on $Nb_3Sn$ superconductors are being studied as the baseline for the main dipoles [1]. Specifications for $Nb_3Sn$ conductors, including a non-Cu $J_c$ of 1500 A/mm$^2$ at 4.2 K and 16 T (corresponding to ~1950 A/mm$^2$ at 4.2 K and 15 T), have been selected and are being used for the magnet designs [2]. Magnets built with conductors of lower $J_c$ would require larger coils, which would lead to a large increase in cost. The $J_c$ of the state-of-the-art $Nb_3Sn$ conductors, which are the rod-restack-process (RRP) type, is not as high as we need: even the record $J_c$ reached by the top-performing RRP conductors is 10-15% below the above target [3,4]; moreover, the ~30% spread of $J_c$ among RRP billets which is seen in wires with subelement size ($D_s$) of 55 μm [3,4] means that in fact a ~45-50% improvement in $J_c$ is required to reach the needs of FCC. After substantial optimization in the past two decades, RRP wires have become excellent technical conductors, and significant achievements have been made in reducing $D_s$ [5], but the $J_c$ performance has been at a plateau since the early 2000s [6]. Extensive efforts have been made to improve the $J_c$ level, including fine tuning of precursors such as local area ratios (LAR), doping, and heat treatments; however, these have not led to any substantial change to the stagnant performance levels.

It is shown in [4] that there are three factors determining the non-Cu $J_c$ of a uniform $Nb_3Sn$ conductor: fine-grain (FG) $Nb_3Sn$ fraction within subelements, irreversibility field ($B_{irr}$), and fluxon pinning capacity; of the three, the greatest potential for further improving non-Cu $J_c$ lies in enhancing fluxon pinning [4]. For example, proton or neutron irradiation on reacted $Nb_3Sn$ conductors [7,8] demonstrated significantly improved non-Cu $J_c$s by introducing point pinning centers (e.g., to the record-breaking 2270 A/mm$^2$ at 4.2 K and 15 T [7]), although using irradiation processing for wind-and-react Nb$_3$Sn magnets is difficult. In fact, introducing additional pinning centers (APC) to Nb$_3$Sn conductors has long been a "holy grail" in the Nb$_3$Sn community, with various efforts on a number of techniques since the 1980s, but none succeeded in making high-$J_c$ wires [9-13]. The internal oxidation (IO) technique, which adds Zr and O to Nb for forming ZrO$_2$ particles that refine Nb$_3$Sn grains, is a promising approach given its big success in the 1960s in Nb$_3$Sn tapes reacted above 950 °C [14]; nevertheless, transferring the technique to wires was challenging. Attempts to apply it in Nb$_3$Sn wires reacted below 800 °C by Zeitlin with various ingenious methods did not show grain refinement [12]. Nearly 10 years later, in a review of the literature, we worked out the essentials for applying the IO technique in Nb$_3$Sn wires [15], and then successfully realized it: we achieved first on binary monofilament wires [15,16] a refinement of average grain sizes from 120-150 nm down to 35-50 nm, as well as a doubling of Nb$_3$Sn layer $J_c$ at 12 T despite its low $B_{irr}$ (20 T) that is mostly due to lack of Ta/Ti dopant. It also showed shift of pinning force versus field ($F_p$-$B$) curve peaks to higher field (from 0.2$B_{irr}$ to 0.33$B_{irr}$), which enhanced high-field $J_c$ and suppressed low-field $J_c$ [16]. With this, engineering fluxon pinning of Nb$_3$Sn wires finally became a reality, leading to great promise to push their performance to much higher levels.

Also in [16] we proposed applying this technique in the simple powder-in-tube (PIT) design where oxide powder is blended into Sn source powders as filament cores. Its feasibility was reinforced by the observation in early 2015 of oxygen transfer from oxide to Nb without direct contact (i.e., via vapor) [17,18]. Then SupraMagentics Inc. fabricated some binary PIT-APC wires with 120 filaments, which were characterized at Florida State University (FSU) and Brookhaven National Lab (BNL) [19], as well as Ohio State University (OSU) [20]: both effects of refined grain size (<40 nm) and shift in $F_p$-$B$ curve peak were clearly verified [19,20]. Hyper Tech Research Inc. (HTR) fabricated some 61-filament binary PIT-APC wires, which were drawn to $D_s$ of 24 µm [18]. However, both SupraMagnetics and HTR's early binary PIT-APC wires had low $B_{irr}$ and needed improvement in wire quality and recipe.

Development of ternary PIT-APC multifilamentary wires began in earnest from 2017, as a collaboration between Fermilab, HTR, and OSU, with focus on improving conductor design, including filament architecture, precursor selection, precursor ratios, tube size, and powder packing density, etc. These efforts have led to significant improvement of performance of the APC conductors in the past two years, while this improvement process is still ongoing. Meanwhile, these progresses also aroused great interest in $Nb_3Sn$ pinning work in other groups recently. In parallel there is also some interesting work on APC $Nb_3Sn$ with new alloys at FSU [21]. Also, exciting results have been obtained on APC $Nb_3Sn$ at University of Geneva based on the IO method [22]. In this letter we demonstrate that ternary PIT-APC $Nb_3Sn$ conductors with high $B_{c2}$ and $B_{irr}$ have been achieved, with competitive transport $J_c$, and also discuss the potential for further $J_c$ improvement.

## 2. Experimental

### 2.1. Samples

The PIT-APC multifilamentary wires used for this work were fabricated by HTR. So far all of the APC wires have been made using a 48/61 design (i.e., 48 $Nb_3Sn$ filaments and 13 Cu hexagonal rods) with a Cu/non-Cu ratio of ~1.3. Designs with higher filament counts are in the preparation stage. All wires were fabricated with a starting billet diameter of 19 mm and drawn to various final wire diameters of 0.5-1.0 mm, resulting in a total length of 100-200 meters for each billet depending on final diameters. After continuous optimization of conductor recipe and improvement of the fabrication process, presently all APC wires can be routinely drawn to the final sizes in single pieces: none of the past 15 billets had any wire breakage.

Two of our most recent multifilamentary APC conductors have been tested at the National High Magnetic Field Laboratory (NHMFL) and the results are reported here. Conductor A (T3851, denoted as APC-A) used Nb-0.6%Zr-3at.%Ta tube, in which both Zr and Ta contents fell short of target due to problems in making that batch of alloy, while Conductor B (T3882, denoted as APC-B) used Nb-1%Zr-4at.%Ta tube. Both conductors used mixture of Sn, Cu, and $SnO_2$ powders. APC-A had enough oxygen for each filament, but the recipe was not yet fully optimized for Cu/Sn and Nb/Sn ratios, as will be discussed below. On the other hand, it was found that APC-B had an insufficient level of oxide due to unoptimized recipe and precursor selection: as a result, some filaments were only partially oxidized.

In order to compare results of the APC wires and present $Nb_3Sn$ conductors, two reference wires were also measured. One is a standard Ti-doped RRP wire for quadrupole magnets in the High Luminosity LHC (HL-LHC) upgrade, which has a diameter of 0.85 mm and 108/127 subelements; other parameters specified in [3]. The other is a standard commercial Ta-doped tube type (TT) wire made by HTR, T1505, with a diameter of 0.7 mm and 192/217 filaments; other parameters described in [23]. The two wires are denoted as "RRP" and "TT", respectively.

All wires were heat treated either in straight segments or on standard ITER barrels under vacuum or, alternatively, in argon-flowing furnaces. APC-A (0.7 mm diameter) was reacted at 675 °C for 152 hours, and APC-B (0.84 mm diameter) was reacted at 675 °C for 300 hours, both with a standard ramp rate of 30 °C/h without any intermediate steps. The RRP wire was heat treated using the standard protocol: 210°C/48h + 400°C/48h + 665°C/75h [3]. The TT wire was heat treated at 625 °C for 400 hours. It is known from various earlier studies [4,24,25] that $B_{irr}$s of PIT and TT conductors are not sensitive to reaction temperature in the 625-675 °C range.

*2.2. Measurements*

Two different types of tests were performed at NHMFL. The first was a measurement of resistance vs field (*R-B*), and was used to determine $B_{irr}$ and $B_{c2}$ as in [26]. The second was voltage vs current (*V-I*) taken at a given *B*. All measurements used a four-point method and were performed at 4.2 K in a 31 T DC magnet at NHMFL with fields perpendicular to wire axes. For each measurement, great care was taken to ensure the samples were centered within the magnet. For the *R-B* measurements, the sample length was 15 mm and the voltage tap separation was 5 mm; a sensing current of 31.6 mA was used. Because 5 samples could be mounted together for each run, we always included at least one reference wire which gave a second check on the results. The voltages were recorded both for increasing and decreasing magnetic fields; hysteresis was minimized by the choice of time constant and was embedded in our error estimates for the critical field determinations (~0.2 T). For *V-I* tests on straight samples, each segment of 35 mm was measured with voltage tap separation of 5 mm, and a criterion of 1 µV/cm was used to determine the critical current ($I_c$). For ITER barrels the voltage tap separation was 50 cm and a criterion of 0.1 µV/cm was used to determine $I_c$.

## 3. Results and Discussions

The *R-B* curves of APC-A and APC-B are shown in Fig. 1, along with RRP and TT. Also included is an APC wire fabricated in 2017 (denoted as "APC17") which used the same Nb-0.6%Zr-3at.%Ta tube as APC-A but had much lower wire quality (e.g., with bad filaments). APC17 at 0.7 mm was reacted at 675 °C for 150 hours. The fields at 10%, 50%, and 90% of the $R$-$B$ transitions are shown in Table 1. Below we take 10% as $B_{irr}$ and 90% as $B_{c2}$.

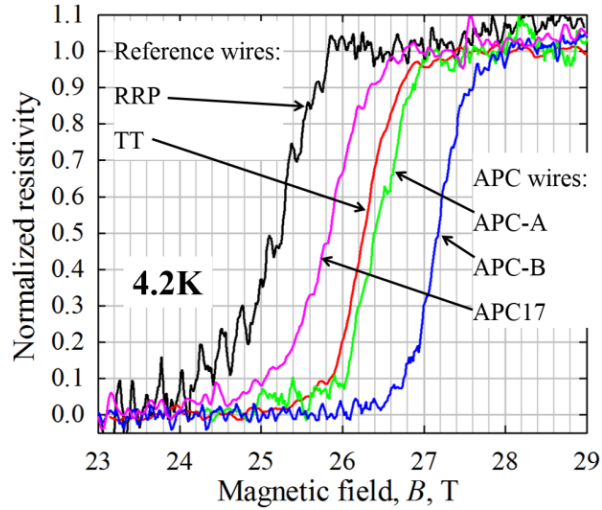

Figure 1. The $R$-$B$ curves of the two reference and three APC wires.

Table 1. $B_{c2}$ (4.2 K) values as determined from the $R$-$B$ curves

|  | RRP | TT | APC-A | APC-B | APC17 |
|---|---|---|---|---|---|
| $B_{c2}$-10% | -* | 25.9 | 26.0 | 26.8 | 25 |
| $B_{c2}$-50% | 25.2 | 26.3 | 26.4 | 27.2 | 25.8 |
| $B_{c2}$-90% | 25.8 | 26.7 | 26.9 | 27.6 | 26.3 |

*: The accurate $B_{c2}$ of RRP at 10% of transition is difficult to define due to noise in the curve.

The $B_{c2}$ of the RRP wire is 25.8 T, which is in the typical range for RRP conductors reacted at 665 °C. The TT wire has $B_{c2}$ of 26.7 T, ~1 T higher than the RRP wire. Higher $B_{c2}$s observed in TT and non-APC PIT conductors as compared to RRP given heat treatments optimal for best $J_c$ performance have been reported by numerous studies [4,24,25]. This is most likely due to better stoichiometry in TT and PIT conductors, in which the Sn source is mostly $Nb_6Sn_5$ and high-Sn Cu-Sn which have high chemical potentials of Sn. A recently-developed theory for stoichiometry of Nb$_3$Sn conductors explains this effect [4,27]. The $B_{c2}$ of APC-A is 26.9 T, ~0.2 T higher than TT. As mentioned earlier, APC-A wire has insufficient Ta content (3 at.%); its $B_{c2}$ could be higher if the Ta content was increased to the optimal level (4 at.%). The APC-B had a $B_{c2}$ of 27.6 T, ~2 T higher than RRP.

The above results clearly show that the IO technique slightly enhances $B_{c2}$ of Nb$_3$Sn instead of reducing it. This is quite different from NbTi conductors, in which a clear suppression of $B_{c2}$ was seen when APCs were incorporated [28]. The reason is that in NbTi conductors APCs were typically normal conducting metals, which suppress superconductivity due to the proximity effect [28]. On the other hand, the IO technique forms ZrO$_2$ particles in Nb$_3$Sn; because ZrO$_2$ is a ceramic and insulator, it has no effect on the superconductivity of Nb$_3$Sn. The low $B_{irr}$s obtained on previous monofilaments [16] and multifilamentary wires [18,19] could be due to lack of Ta or Ti dopant and poor wire quality. Indeed, wire quality has a significant influence on $B_{c2}$: for example, APC17 has noticeably lower $B_{c2}$ than APC-A and even TT. The APC-A and APC-B wires are better in quality and thus exhibit higher $B_{c2}$s.

The non-Cu $J_c$s calculated from the $I_c$s of APC-A and APC-B are shown in Fig. 2, along with the reference RRP wire. The data points are fitted with the commonly-used two-component pinning force model [11], which has been proven by several studies to work well for APC conductors [7,11,19]. As can be seen from Fig. 2, the fittings to the data points are very good. The obtained $B_{irr}$ parameters from the fittings are 24.6 T, 25.2 T, and 26.2 T for RRP, APC-A, and APC-B, respectively. These values agree well with the fields where transitions of the $R$-$B$ curves start, further confirming the $B_{irr}$ results. The non-Cu $J_c$s of RRP are ~1090 and ~2680 A/mm$^2$ at 16 T and 12 T, respectively, which are not the highest values reached by the top-performing RRP conductors, but still in the typical $J_c$ spectrum for a RRP wire with $D_s$ of 55 μm. The non-Cu $J_c$s at 16 T of APC-A and APC-B are 1150 and 1040 A/mm$^2$, respectively. Both APC wires had clear advantage in high-field $J_c$s relative to RRP due to higher $B_{irr}$s and shift in $F_p$-$B$ curve peaks.

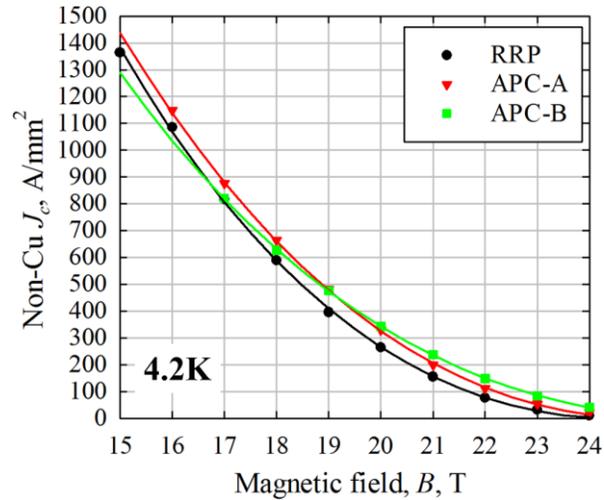

Figure 2. Non-Cu $J_c$s (4.2 K) of RRP, APC-A, APC-B.

Back-scattered electron (BSE) images of polished cross sections of the two APC wires are shown in Fig. 3. From the images it can be calculated that the average FG area fractions in filaments are ~33% and ~22% for APC-A and APC-B, respectively, and the residual Nb area fractions are 36% and 45%, respectively. The calculated coarse-grain (CG) fractions for APC-A and APC-B are ~13%, similar to those of TT [25] and PIT [29] wires. In comparison, the residual Nb fractions for present PIT conductors are typically below 25% [29]. This means that the above APC conductors can form more FG by converting more Nb via properly tweaking the recipe design (e.g., Nb/Sn ratio) and optimizing reaction so that no Sn remains in the core. Our calculation shows that by doing the above and reducing the barrier fraction to 25%, the FG fraction would reach 40% or slightly above, as in TT and PIT conductors [25,29].

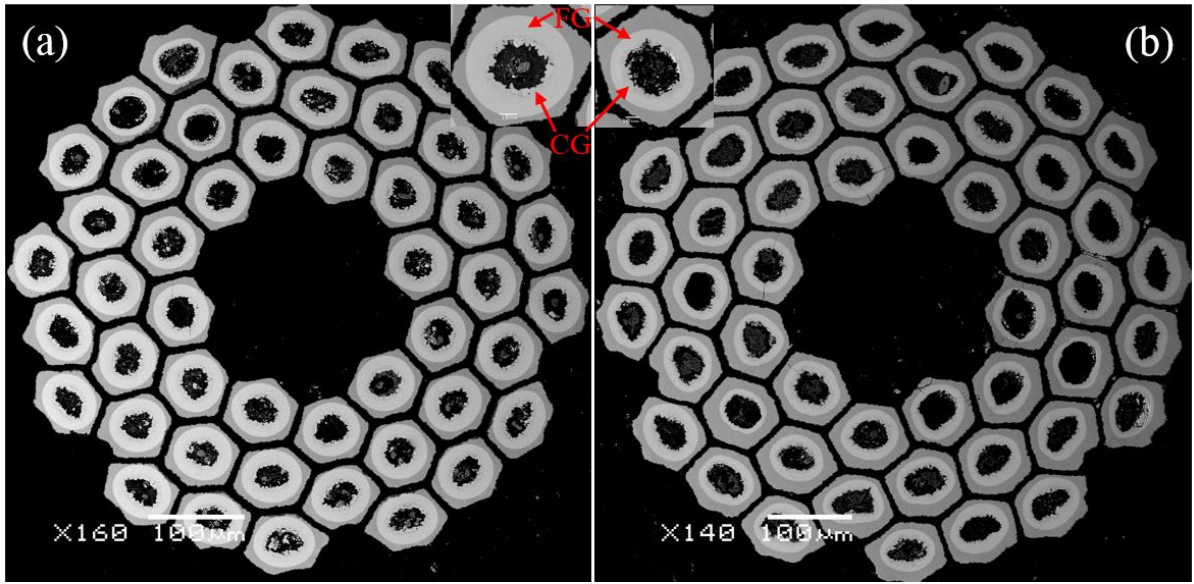

Figure 3. BSE images of cross sections for reacted (a) APC-A and (b) APC-B.

Also seen from Fig. 3 is that there is still some variation of $Nb_3Sn$ fraction among filaments, which is also seen along the wire length. This apparently suppresses the overall $I_c$ because the maximum current carried by a filament is determined by its worst segment. Such a non-uniformity issue may be caused by inhomogeneous powder distribution or perhaps some local Sn leakage into the Cu matrix before filaments are fully reacted. The former issue can be solved by improving powder mixing and filling techniques. For the latter case, the most common issue is local thinning of Nb alloy tube (e.g., filament cores being off the centers). This issue has been largely solved by development work in the past two years via proper attention to powder precursors as well as their ratios and mixing. Further improvements are being pursued.

From the above measured non-Cu $J_c$s and FG area fractions, the Nb$_3$Sn layer $J_c$s are calculated and shown in Fig. 4. The 16 T layer $J_c$s for RRP, APC-A, APC-B are 1850, 3450 and 4710 A/mm$^2$, respectively, while that of the top-performing RRP wires is ~2200 A/mm$^2$. Clearly the APC wires have higher advantage at higher fields.

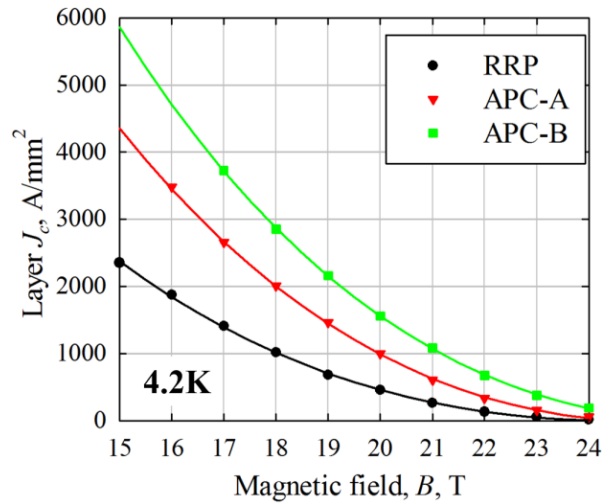

Figure 4. Nb$_3$Sn layer $J_c$s (4.2 K) of RRP, APC-A, APC-B.

Secondary-electron (SE) images of fractured cross sections of the two APC samples are shown in Fig. 5. It was found that grain size distributions in the FG layers are not uniform. The calculated average grain sizes of APC-A and APC-B after 675°C reactions were 81 and 72 nm, respectively, which are about half of those RRP and PIT wires reacted at 650-675 °C [4,30], but much larger than the average grain sizes of the early APC samples with 1% Zr and sufficient O reacted below 650 °C, which were 35-45 nm [15,19,20]. For APC-A, this is because it was reacted at 675 °C and it had only 0.6%Zr in the Nb alloy: lower Zr content leads to fewer ZrO$_2$ particles for grain refinement. For APC-B, the less refined grain size is because it was reacted at 675 °C and it did not have sufficient oxygen due to an unoptimized recipe. Earlier studies indicate that grain size is sensitive to oxygen content, and that by optimizing the conductor recipe to fully oxidize all filaments, the average grain size can be reduced to 50-65 nm for reactions at 650-675 °C [15,18]. With $B_{irr}$ above 25 T, grain sizes of 50-65 nm are expected to lead to a Nb$_3$Sn layer $J_c$ at 16 T of 5000-6000 A/mm$^2$, as a conservative estimation based on the relation between $J_c$ and grain size [16,30].

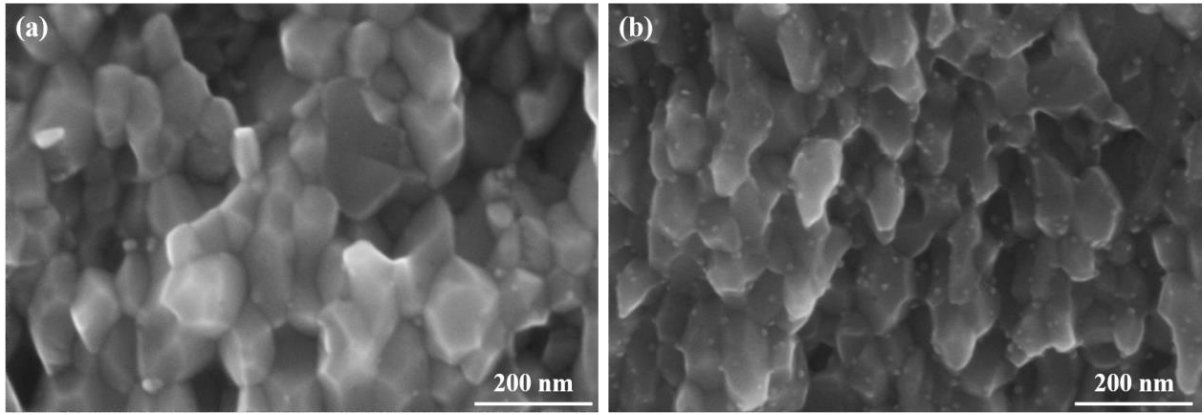

Figure 5. SE images of fractured cross sections of (a) APC-A (average grain size 81 nm) and (b) APC-B (average grain size 72 nm).

## 4. Conclusions

Although the APC technique was first realized in Nb$_3$Sn wires in 2014, focused development of ternary multi-filamentary PIT-APC wires began only in 2017, and since then progress has been fast. Measurements at NHMFL of the recently-developed ternary APC conductors gave $B_{irr}$s and $B_{c2}$s in the 26-28 T range, values that are at or above those of the best RRP conductors. The non-Cu $J_c$s of these APC conductors have reached nearly 1200 A/mm$^2$ at 16 T and 4.2 K, similar to present RRP conductors, in spite of the fact that the FG fractions are still significantly below those of present-day PIT, TT, and RRP conductors due to the lack of recipe and heat treatment optimizations. The measured Nb$_3$Sn layer $J_c$ of the APC wire with 1%Zr reacted at 675 °C is 4710 A/mm$^2$ at 16 T and 4.2 K, while that of RRP is below 2000 A/mm$^2$. The reason for the large increase in layer $J_c$ for the APC conductor is mainly enhanced fluxon pinning, due to refined grain size and also ZrO$_2$ APCs in Nb$_3$Sn grains. The APC wire with 1%Zr had average grain size of ~72 nm, about half of those of RRP conductors, but can still be further reduced via recipe and heat treatment optimizations. We are presently working to optimize the conductor recipe and heat treatment in order to fully react the filaments (targeting 40% FG), as well as reduce the grain size to the previously observed 50-65 nm by fully oxidizing all of the filaments. If the targeted FG fraction and layer $J_c$ can be reached, the non-Cu $J_c$ at 16 T and 4.2 K should scale to 2000 A/mm$^2$ or above, greater than the target for FCC so that sufficient margin can be provided, representing a crucial advance for Nb$_3$Sn conductors.

**Acknowledgements**

This work was supported by the Laboratory Directed Research and Development (LDRD) program of Fermi National Accelerator Laboratory and HTR SBIR DE-SC0013849 and DE-SC0017755 by US Department of Energy. A portion of this work was performed at the NHMFL, which is supported by National Science Foundation Cooperative Agreement No. DMR-1644779 and the State of Florida. Some measurements were conducted in the Center for Nanoscale Materials, Argonne National Laboratory, which was supported by the U.S. Department of Energy, Office of Science, Office of Basic Energy Sciences, under Contract No. DE-AC02-06CH11357. The measurements at the NHMFL were greatly helped by Jan

Jaroszynski, Griffin Bradford, and Yavuz Oz from FSU. We thank Ian Pong from LBNL for supplying the RRP wire that is used as a reference in this work. We acknowledge some discussions under the U.S. Magnet Development Program (MDP) context.**References**

[1]. Benedikt M and Zimmermann F 2016 Status of the future circular collider study *CERN-ACC-2016-0331*
[2]. Tommasini D and Toral F 2016 Overview of magnet design options *EuroCirCol-P1-WP5 report* 4-6
[3]. Cooley L D, Ghosh A K, Dietderich D R and Pong I 2017 Conductor Specification and Validation for High-Luminosity LHC Quadrupole Magnets *IEEE Trans. Appl. Supercond.* **27** 6000505
[4]. Xu X 2017 A review and prospects for $Nb_3Sn$ superconductor development *Supercond. Sci. Technol.* **30** 093001
[5]. Parrell J A *et al.* 2009 Internal tin $Nb_3Sn$ conductors engineered for fusion and particle accelerator applications *IEEE Trans. Appl. Supercond.* **19** 2573-9
[6]. Parrell J A, Zhang Y Z, Field M B, Cisek P and Hong S 2003 High field $Nb_3Sn$ conductor development at Oxford Superconducting Technology *IEEE Trans. Appl. Supercond.* **13** 3470-3
[7]. Baumgartner T, Eisterer M, Weber H W, Flükiger R, Scheuerlein C and Bottura L 2015 Performance Boost in Industrial Multifilamentary $Nb_3Sn$ Wires due to Radiation Induced Pinning Centers *Sci. Rep.* **5** 10236
[8]. Spina T, Scheuerlein C, Richter D, Bordini B, Bottura L, Ballarino and Flükiger R 2015 Artificial Pinning in Nb3Sn Wires *IEEE Trans. Appl. Supercond.* **27** 8001205
[9]. Flukiger R, Specking W, Klemm M and Gauss S 1989 Composite core $Nb_3Sn$ wires: preparation and characterization *IEEE Trans. Magn.* **25** 2192-9
[10]. Zhou R, Hong S, Marancik W and Kear B 1993 Artificial flux pinning in Nb and $Nb_3Sn$ superconductors *IEEE Trans. Appl. Supercond.* **3** 986-9
[11]. Rodrigues D, Da Silva L B S, Rodrigues C A, Oliveira N F and Bormio-Nunes C 2011 Optimization of Heat Treatment Profiles Applied to Nanometric-Scale $Nb_3Sn$ Wires With Cu-Sn Artificial Pinning Centers *IEEE Trans. Appl. Supercon.* **21** 3150-3
[12]. Zeitlin B A, Gregory E, Marte J, Benz M, Pyon T, Scanlan R and Dietderich D 2005 Results on Mono Element Internal Tin $Nb_3Sn$ Conductors (MEIT) with Nb7.5Ta and $Nb(1Zr+O_x)$ Filaments *IEEE Trans. Appl. Supercond.* **15** 3393-6
[13]. Motowidlo L R, Lee P J and Larbalestier D C 2009 The Effect of Second Phase Additions on the Microstructure and Bulk Pinning Force in $Nb_3Sn$ PIT Wire *IEEE Trans. Appl. Supercond.* **19** 2568-72
14

[14]. Rumaner L E, Benz M G and Hall E L 1994 The role of oxygen and zirconium in the formation and growth of $Nb_3Sn$ grains *Metallurgical and Materials Transactions A* **25** 213-9
[15]. Xu X, Sumption M D, Peng X and Collings E W 2014 Refinement of $Nb_3Sn$ grain size by the generation of $ZrO_2$ precipitates in $Nb_3Sn$ wires *Appl. Phys. Lett.* **104** 082602
[16]. Xu X, Sumption M D and Peng X 2015 Internally Oxidized $Nb_3Sn$ Superconductor with Very Fine Grain Size and High Critical Current Density *Adv. Mater.* **27** 1346-50
[17]. Xu X, Sumption M D and Peng X 2018 Superconducting wires and methods of making thereof *U.S. Patent 9916919*, filed date: Feb. 18, 2015
[18]. Xu X, Peng X, Sumption M D and Collings E W 2017 Recent Progress in Application of Internal Oxidation Technique in $Nb_3Sn$ Strands *IEEE Trans. Appl. Supercond.* **27** 6000105
[19]. Motowidlo L, Lee P J, Tarantini C, Balachandran S, Ghosh A and Larbalestier D C 2017 An intermetallic powder-in-tube approach to increased flux-pinning in $Nb_3Sn$ by internal oxidation of Zr *Supercond. Sci. Technol.* **31** 014002
[20]. Unpublished data
[21]. Balachandran S, Tarantini C, Lee P J, Kametani F, Su Y, Walker B, Starch W L and Larbalestier D C Beneficial influence of Hf and Zr additions to Nb4at.%Ta on the vortex pinning of Nb3Sn with and without an O Source *https://arxiv.org/abs/1811.08867*
[22]. Buta F, Bonura M, Matera D, Ballarino A, Hopkins S, Bordini B and Senatore C 2018 Properties and microstructure of binary and ternary $Nb_3Sn$ superconductors with internally oxidized $ZrO_2$ nanoparticles *Applied Supercond. Conf.* 1MPo2A-06
[23]. Xu X, Collings E W, Sumption M D, Kovacs C and Peng X 2014 The effects of Ti addition and high Cu/Sn ratio on tube type $(Nb,Ta)_3Sn$ strands, and a new type of strand designed to reduce unreacted Nb ratio *IEEE Trans. Appl. Supercond.* **24** 6000904
[24]. Boutboul T, Oberli L, Ouden A, Pedrini D, Seeber B and Volpini G 2009 Heat Treatment Optimization Studies on PIT $Nb_3Sn$ Strand for the NED Project *IEEE Trans. Appl. Supercond.* **19** 2564-7
[25]. Xu X, Sumption M D, Bhartiya S, Peng X and Collings E W 2013 Critical current densities and microstructures in rod-in-tube and Tube Type $Nb_3Sn$ strands – Present status and prospects for improvement *Supercond. Sci. Technol.* **26** 075015
[26]. Godeke A, Jewell M, Fischer C, Squitieri A, Lee P and Larbalestier D 2005 The upper critical field of filamentary $Nb_3Sn$ conductors *J. Appl. Phys.* **97** 093909
[27]. Xu X and Sumption M 2016 A model for the compositions of non-stoichiometric intermediate phases formed by diffusion reactions, and its application to $Nb_3Sn$ superconductors *Sci. Rep.* **6** 19096
[28]. Motowidlo L R, Zeitlin B A, Walker M S and Haldar P 1992 Multifilament NbTi with artificial pinning centers: The effect of alloy and pin material on the superconducting properties *Appl. Phys. Lett.* **61** 991-4
[29]. Segal C, Tarantini C, Lee P J and Larbalestier D C 2017 Improvement of small to large grain A15 ratio in $Nb_3Sn$ PIT wires by inverted multistage heat treatments *IOP Conf. Ser. Mater. Sci. Eng.* **279** 012019
[30]. Xu X, Sumption M D and Collings E W 2014 Influence of Heat Treatment Temperature and Ti doping on Low Field Flux Jumping and Stability in $(Nb-Ta)_3Sn$ Strands *Supercond. Sci. Technol.* **27** 095009